# Epitaxial graphene homogeneity and quantum Hall effect in millimeter-scale devices

*Yanfei Yang[†,\*], Guangjun Cheng[†], Patrick Mende[‡], Irene G. Calizo[†,§], Randall M. Feenstra[‡], Chiashain Chuang[†,Π], Chieh-Wen Liu[†,£], Chieh-I Liu[†,£], George R. Jones[†], Angela R. Hight Walker[†], and Randolph E. Elmquist[†]*

[†]National Institute of Standards and Technology (NIST), Gaithersburg, MD 20899-8171, USA

[‡]Department of Physics, Carnegie Mellon University, Pittsburgh, PA 15213-3890, USA

[§] Present address: Mechanical and Materials Engineering, Florida International University, Miami, FL 33174

[Π] Present address: Graduate School of Advanced Integration Science, Chiba University, Chiba, 263-8522, Japan

[£]Graduate Institute of Applied Physics, National Taiwan University, Taipei 106, Taiwan

*E-mail: yanfei.yang@nist.gov

ABSTRACT: Quantized magnetotransport is observed in 5.6 × 5.6 mm$^2$ epitaxial graphene devices, grown using highly constrained sublimation on the Si-face of SiC(0001) at high temperature (1900 °C). The precise quantized Hall resistance of $R_{xy} = \frac{h}{2e^2}$ is maintained up to record level of critical current $I_{xx} = 0.72$ mA at $T = 3.1$ K and 9 T in a device where Raman microscopy reveals low and homogeneous strain. Adsorption-induced molecular doping in a second device reduced the carrier concentration close to the Dirac point ($n \approx 10^{10}$ cm$^{-2}$), where mobility of 43 700 cm$^2$/Vs is measured over an area of 10 mm$^2$. Atomic force, confocal optical, and Raman microscopies are used to characterize the large-scale devices, and reveal improved SiC terrace topography and the structure of the graphene layer. Our results show that the structural uniformity of epitaxial graphene produced by face-to-graphite processing contributes to millimeter-scale transport homogeneity, and will prove useful for scientific and commercial applications.

Wafer-scale monolayer graphene[1,2] can be produced by thermal decomposition of certain polytypes of silicon carbide[3] (SiC) or by chemical vapor deposition (CVD) on metal catalyst substrates[2]. While CVD graphene forms randomly oriented domains to match the crystal orientation of the metal catalyst, epitaxial graphene (EG) forms a single domain on monocrystalline wafers of hexagonal SiC(0001)[4] and the insulating SiC substrate is immediately suitable for fabrication of electronic[5], plasmonic[6] and photonic[7] devices. Quantum Hall effect (QHE) standards produced from EG[8-9] can be operated economically at lower magnetic fields and higher temperatures than GaAs-AlGaAs heterostructures[10]; thus EG devices are likely to become the premier source of resistance traceability in practical metrology and their optimization is of great interest to the electrical metrology community. Here, we report precision measurements of the QHE in millimeter-scale EG devices at high current and correlate the quantized magnetotransport to microscopy data, including structural reorganization of the SiC surface, EG layer number and distribution, and strain as measured by Raman microscopy.

Efforts to produce nearly defect-free monolayer EG on SiC generally involve control of the high-temperature vapor phase. For example, annealing in atmospheric-pressure Ar gas[11] or in a small confining enclosure[12] helps to raise the partial pressures of sublimated Si, $Si_2C$ and $SiC_2$ closer to equilibrium at high temperature, and the number of defects in graphene is then reduced and the morphology of vicinal SiC(0001) surfaces is generally improved. However, dissociated carbon atoms may diffuse anisotropically[13], leading to the formation of multiple graphene layers near the edges of the terraces[14]. Furthermore, SiC restructuring[3] and energetically-favorable step-bunching also may produce undesirable terrace facet edges[14,15] that face off-axis by $\approx 30°$ on vicinal SiC(0001). Atomic-layer-resolved characterization has shown significant delamination of the carbon buffer layer[15,16] on facet edges that separate adjacent terraces.

To minimize these possible complications due to the substrate, we produce EG using a constraint on vapor diffusion provided by close proximity to polished pyrolytic graphite (see Fig. 1a). This face-to-graphite[17] (FTG) method leads to uniform EG growth with limited terrace restructuring on clean, low-miscut, chemically-mechanically polished SiC(0001) substrates[18], and often results in crescent-shaped terraces having small areas and low aspect ratios, as shown in Fig. 1b. Figure 1c shows details of the topographic structure produced by the FTG process at 1900 °C on the surface of low-miscut SiC(0001), imaged by atomic force microscopy (AFM), and the corresponding phase image (Fig. 1d) shows uniform contrast. We have found that this terrace topography, together with the near-equilibrium FTG environment, supports more isotropic carbon diffusion compared to parallel, linear terraces, thus reducing the tendency to form extended bilayer ribbons. Annealing samples at $T > 1800$ °C with two SiC(0001) surfaces arranged face-to-face[18,19] results in uncontrolled step-bunching of the terraces (see Supplement Fig. S1a,b). This large change in substrate topography may be produced by the vapor phase transfer of molecules between the facing samples. Under similar conditions, but annealing the sample facing Ar background gas, it has been shown elsewhere that sub-μm scale pits are likely to develop on low-miscut SiC substrates[20] (see Supplement Fig. S1c). Our results show that FTG confinement reduces structural disorder due to pitting of the SiC surface for low-miscut substrates. Reduced mobility in EG transport has been correlated with the frequency of pitting[20] and with the size and height of underlying SiC terraces[21,22].

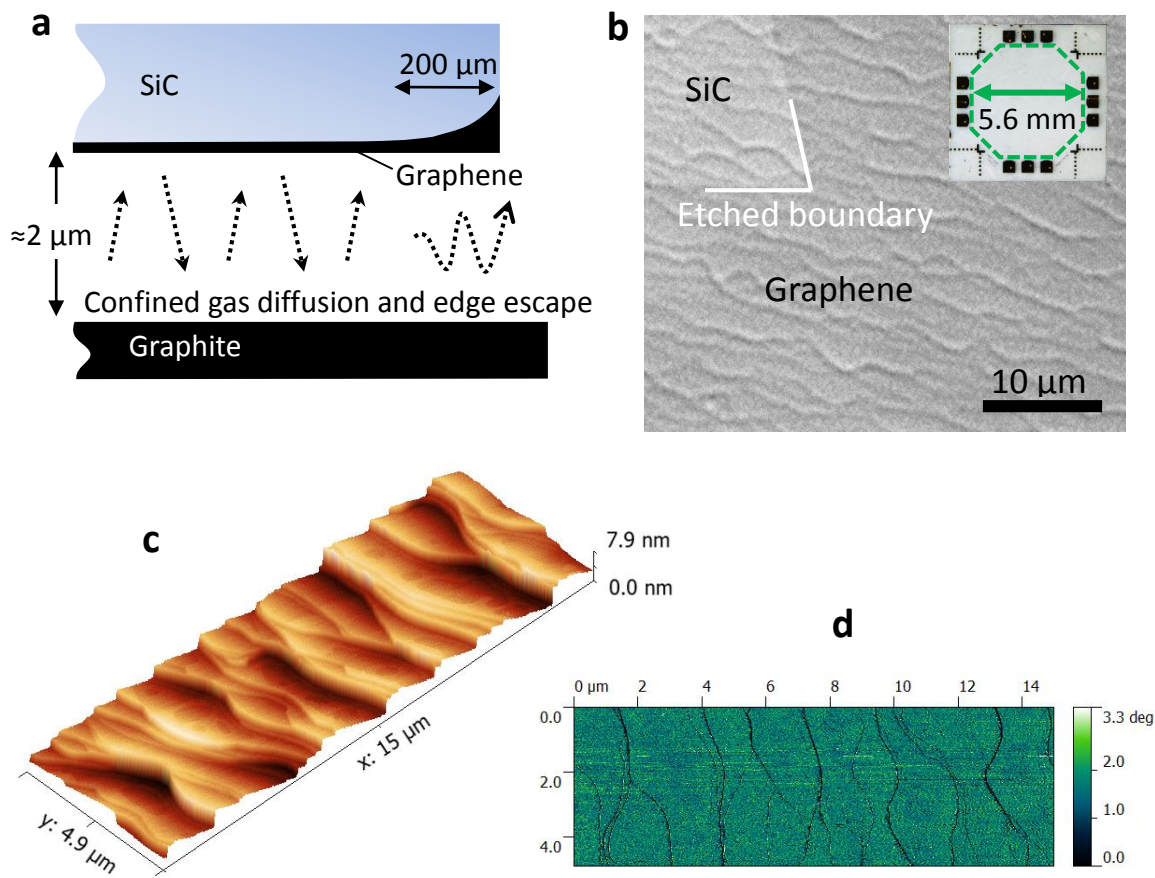

**Figure 1.** Millimeter-scale EG growth by FTG method. (a) Diagram showing a cross-section of the SiC sample and polished graphite during FTG processing. (b) Optical image of a region of sample A, with enhanced contrast to show where EG has been removed (upper left of image), labeled SiC. Inset shows a large-scale device with twelve symmetric gold contacts. (c) AFM

topography image of a FTG sample with step height displayed as contours in the vertical direction. (d) AFM phase image corresponding to height image (c).

Inhomogeneous transport characteristics may result from substrate topography, variation in EG layer number[23-26], and/or the effect of strain[26-33]. We correlate device transport in our FTG-grown EG with layer and strain homogeneity. Raman spectroscopy is a powerful nondestructive technique used to characterize atomically thin graphene samples. The two-phonon G' band of monolayer graphene can be accurately fit by a single Lorentzian. Furthermore, the position of the G' band has been correlated to the strain in the conductive EG lattice[26-28]. Studies of graphene on SiC also have established that the G' band has only a weak dependence on carrier density[29], that strain can change suddenly where the EG layer crosses a terrace edge[30], and that low and uniform strain is often related to improved transport[31].

Deconvolution of the effects of strain on transport in some EG samples can be problematic, as shown in our earlier work where EG was grown at a lower temperature of 1630 °C in Ar background without FTG on a substrate having a high miscut of ≈ 1.26° relative to the SiC(0001) basal plane. Low-energy electron microscopy (LEEM) data confirmed monolayer graphene on step-bunched, parallel terraces of 0.5 μm – 2 μm width, and Raman mapping provided correlated data covering the same region (see Supplement Fig. S2). The G' Raman band was fit to a Lorentzian function resulting in an average G' peak position $\omega_{G'}$ = (2747.4 ± 1.7) cm$^{-1}$ for the sampled region of diameter 44 μm. The low standard deviation of this value indicated excellent homogeneity at the spatial resolution (~2 μm) under which the data was collected. However, the G' linewidth (full-width at half-maximum, FWHM) of $\Gamma_{G'}$ = (63.8 ± 2.6) cm$^{-1}$ was quite broad compared to that seen in graphene produced by exfoliation from graphite[32-33]. Magnetotransport measurements in small Hall bar devices made from this sample revealed low mobilities of $\mu <$

1000 cm$^2$/Vs at 1.5 K. While neither LEEM nor Raman captures the effects of terrace edges, the low mobility in this material may be linked to the broad, albeit uniform, G' linewidth. The large linewidth could result from inhomogeneous strain at sub-micron scale[32] that is averaged within the 1 μm probe volume.

Our present work on devices composed of FTG-grown material shows that the graphene lattice strain and specifically homogeneous low strain is predictive toward EG device transport characteristics at large scale. Exceptional magnetotransport at millimeter-scale was measured in two samples produced using FTG confinement at 1900 °C for 210 s and 235 s, respectively (sample A and sample B). A magnetotransport device of 5.6 mm in height and width was fabricated on each sample, as shown in the inset of Fig. 1b. The surfaces of both samples were characterized by contrast-enhanced optical imaging[22] (Fig. 1b and Supplement Fig. S4), AFM (Figs. 1c,d), and confocal optical microscopy (Figs. 2a,b). On sample A, these imaging techniques show a nearly uninterrupted monolayer with only a few multilayer inclusions, which appear as small irregularly-shaped bright spots or ribbons in reflective confocal microscope images (see Fig. 2a and Supplement Figs. S1d,e). On sample B, bilayer ribbons and small patches of buffer layer are more common in some regions (Fig. 2b), but still occupy a small percentage of the sample surface. Sample B has low terrace topography, but the terraces are wider and more irregular in some areas (Supplement Fig. S4) indicating isolated step bunching. While C-face SiC processed with a graphite cap is reported to develop long graphene ribbons[17], we find that both terrace step bunching and multilayer EG growth are much more limited by FTG growth on the Si-face under optimized growth conditions.

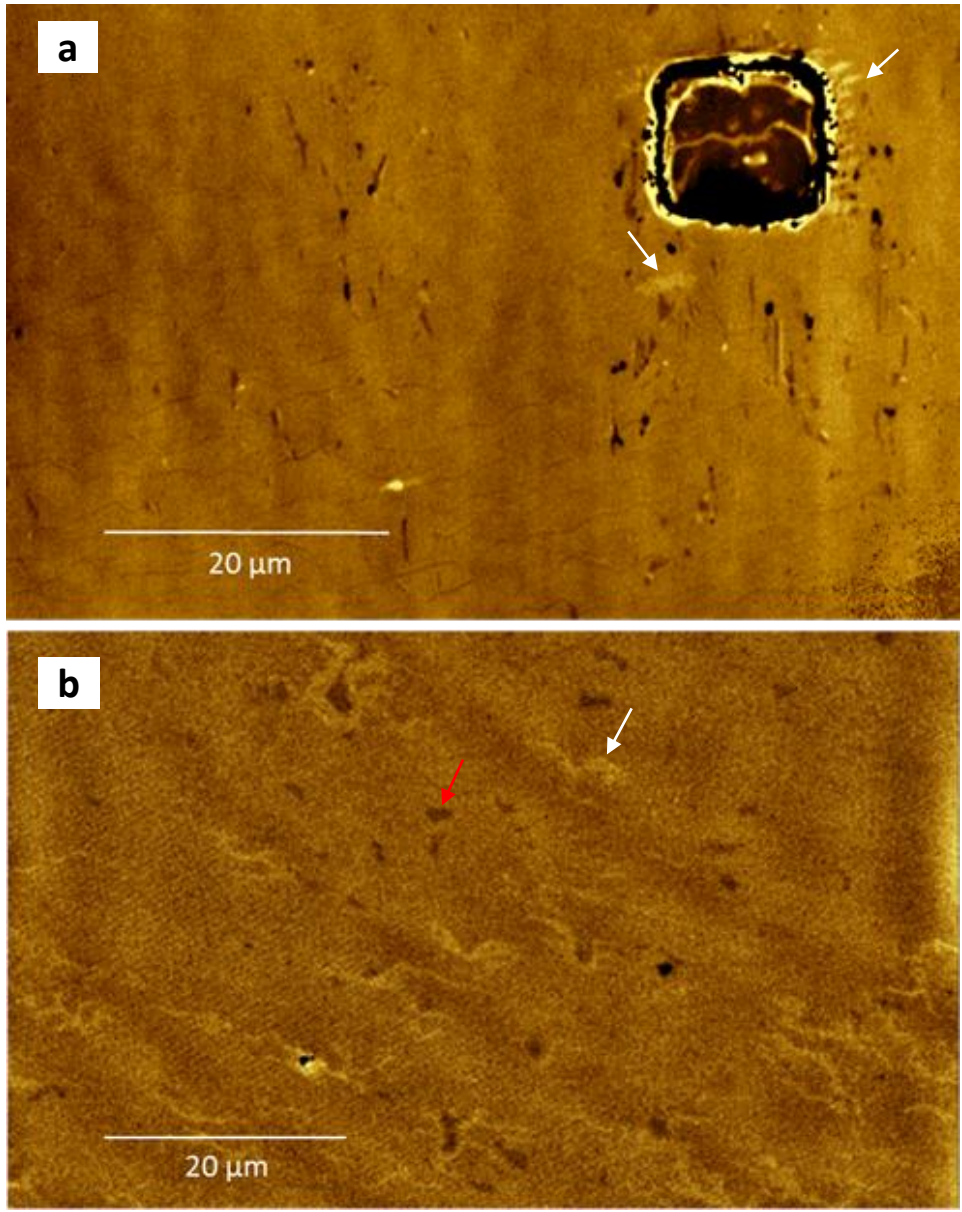

**Figure 2.** Confocal microscopy images of sample A and sample B. (a) Area of sample A, imaged by contrast-enhanced confocal microscopy with reflected 405 nm light. Monolayer graphene appears as a uniform background with terraces barely visible. The square region at upper right is a 500 nm deep etched fiducial mark, showing small fingers of multilayer graphene along its right side, and a bilayer patch just below its lower left corner (indicated by the white arrows). (b) Area of sample B near the left-center contact, imaged by contrast-enhanced confocal microscope with reflected 460 nm light. Monolayer appears as a uniform background, with buffer layer or no graphene showing as darker patches (indicated by the red arrow) and bilayer or multilayer (indicated by the white arrow) appearing as lighter patches or ribbons.

Raman data were obtained in closely-spaced grids after transport devices were fabricated on the two samples. Raman spectra in the D band region near 1350 cm$^{-1}$ at two laser frequencies (Supplement Fig. S3) indicate low-defect EG in both samples. Raman maps (see Methods) were generated for three large, well-separated device regions spanning the midline of sample A, across an area of ≈ 5 mm width. Cross-correlated data for the resulting spectra are shown in Fig. 3a. Of the Raman data collected from the three areas, nearly 99% of the G' spectra (1178 points) are symmetric and can be fit with a single Lorentzian. The fitted G' linewidths are less than 40 cm$^{-1}$, ranging from 27.5 cm$^{-1}$ to 38.2 cm$^{-1}$, and mean center position (2729.7 ± 2.7) cm$^{-1}$, ranging from 2721.3 cm$^{-1}$ to 2737.5 cm$^{-1}$. These closely grouped results with narrow linewidths reveal an unprecedented uniformity for EG, since both strain and layer number variation can affect the fitted parameters. The mean linewidth value $\Gamma_{G'}$ = (31.7 ± 1.4) cm$^{-1}$ may be compared to $\Gamma_{G'}$ ≈ 25 cm$^{-1}$, ranging from 22 cm$^{-1}$ to 35 cm$^{-1}$, as reported recently[33] for exfoliated graphene on SiO$_2$ capped by hexagonal BN (h-BN).

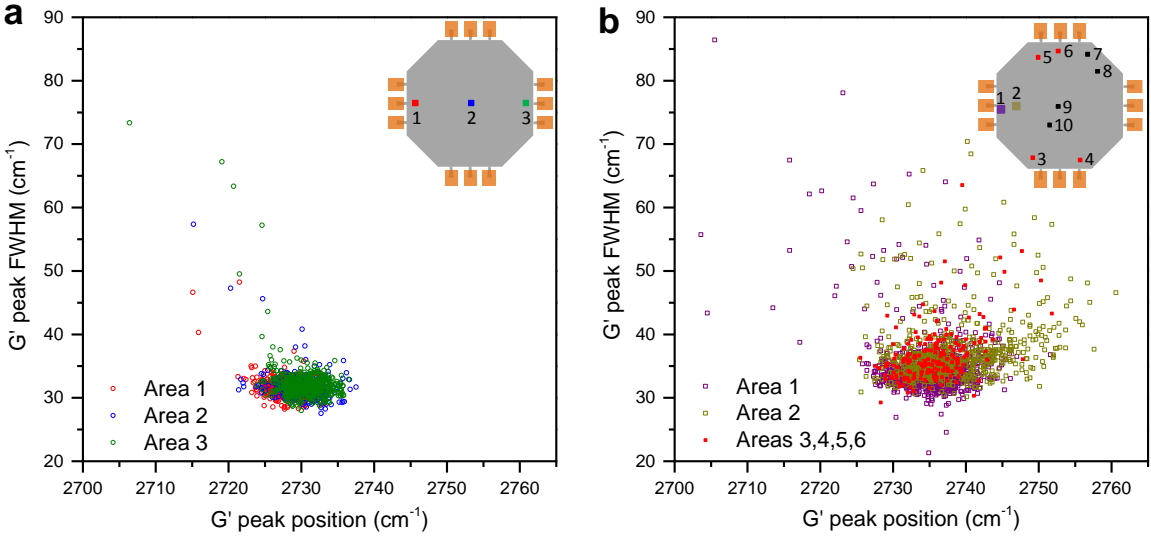

**Figure 3.** Raman microscopy from two 5.6 mm octagonal devices. (a) Correlation between FWHM and peak position for G' peaks collected from three areas of sample A, located as shown in the inset using red, blue, and green. (b) Correlation between FWHM and peak position for G' peaks collected from six areas of sample B, located and identified by color as shown in the inset.

The Raman data from sample B (see Fig. 3b) in general have wider distribution, tending toward higher values of $\Gamma_{G'}$ and $\omega_{G'}$. On the left side of the device we obtained a total of 1718 Raman spectra in two large areas ≈ 1 mm apart near the midline (areas 1 and 2 as seen in the inset of Fig. 3b). These data yield $\Gamma_{G'} = (34.7 \pm 4.6)$ cm$^{-1}$ and $\Gamma_{G'} = (37.3 \pm 5.2)$ cm$^{-1}$, with center positions $\omega_{G'} = (2735 \pm 4.3)$ cm$^{-1}$ and $\omega_{G'} = (2738 \pm 6.3)$ cm$^{-1}$, respectively for area 1 (purple) and area 2 (green). Additional Raman data were collected from eight widely distributed small areas (with less than 64 spectra per area) as shown in the inset of Fig. 3b (areas 3 to 10). While half of these areas (indicated in red) of sample B gave very homogeneous results, falling between the values for both $\omega_{G'}$ and $\Gamma_{G'}$ from the two large areas, the other half indicated in black show no significant overlap (see Supplement Table S1, S2), and gave mostly values of $\Gamma_{G'}$ greater than 40 cm$^{-1}$. We have correlated this inhomogeneity observed in the Raman data from sample B with the positions of bilayer ribbons and buffer layer patches captured in confocal microscopy images, and this will be explored in detail in a subsequent report. To summarize, in sample A we obtain narrow Raman G' linewidths with very little spread in the position, and in sample B we find mostly similar results but with inhomogeneous regions indicated by the Raman data. Neither sample shows linewidths as broad as were seen for the EG sample made by our earlier synthesis method (Fig. S2f).

Electrical measurements on samples A and B were made in a pumped liquid helium cryostat, and will be correlated to our optical measurements. Low-precision AC measurements were used to calculate the carrier density and mobility of the devices, and were repeated at various temperatures and current levels. In fabricating the magnetotransport devices, the active EG surfaces were kept uniformly resist-free by depositing Pd/Au as a thin layer prior to standard photolithography processing, and afterwards removing the Pd/Au layer from the EG region by immersing the devices in dilute aqua regia (by volume, $HNO_3$:$HCl$:$H_2O$ = 1:3:4) for 45 seconds. This fabrication process[34] initiates the attachment of molecular dopants, and can result in carrier concentrations below $n \approx 10^{10}$ cm$^{-2}$ in ungated EG samples, compared to as-grown monolayer EG with substrate-induced doping as high as $n \approx 10^{13}$ cm$^{-2}$. The carrier concentration can be tuned by adding or removing molecular dopant using chemical- or heat-treatment; however, we obtained precise QHE transport results for sample A with the original doping level obtained after removal of the Pd/Au protective layer, at $n \approx 2.3 \times 10^{11}$ cm$^{-2}$ ($\mu \approx 5800$ cm$^2$/Vs). Fig. 4a shows magnetotransport characteristics for sample A at cryogenic temperatures, with the behavior at higher $T$ plotted in the inset.

The hallmarks of the QHE in low carrier density EG are a broad plateau in the Hall resistance $R_{xy}$ with conventional value $\frac{h}{2e^2}$ = 12 906.4035 Ω (see Fig. 4a) and near-zero longitudinal resistivity $\rho_{xx}$. The strength of this QHE plateau at high current and temperature is enhanced at low perpendicular magnetic field strengths by the √$B$ dependence of the Landau level energies, and by field-dependent charge transfer from donors in the SiC substrate and doping layer[8]. The presently used, commercial GaAs devices are rarely capable of sustaining precise QHE measurements at currents above 0.2 mA, and higher currents can exceed the range of most state-of-the-art metrological instruments used to measure $R_{xy}$. For sample A, we used two methods to obtain sensitive characterization of QHE device performance for higher current levels, with the first based on the increase in longitudinal resistivity $\rho_{xx}$. Values of the longitudinal resistivity $\rho_{xx}$ were measured at four temperatures between 1.6 K and 4.2 K over a wide range of source-drain current (0.116 mA to 0.72 mA), as shown in Fig. 4b. With an applied field of $B$ = 9 T, zero dissipation is observed at 1.6 K for the full range of current up to $I_{xx}$ ≈ 0.72 mA, and possibly at higher currents, but this could not be verified with the present apparatus. (Supplement Fig. S5a). These results exceed the highest critical currents reported to date in graphene[9] (0.5 mA) or GaAs heterostructures[35] (0.6 mA).

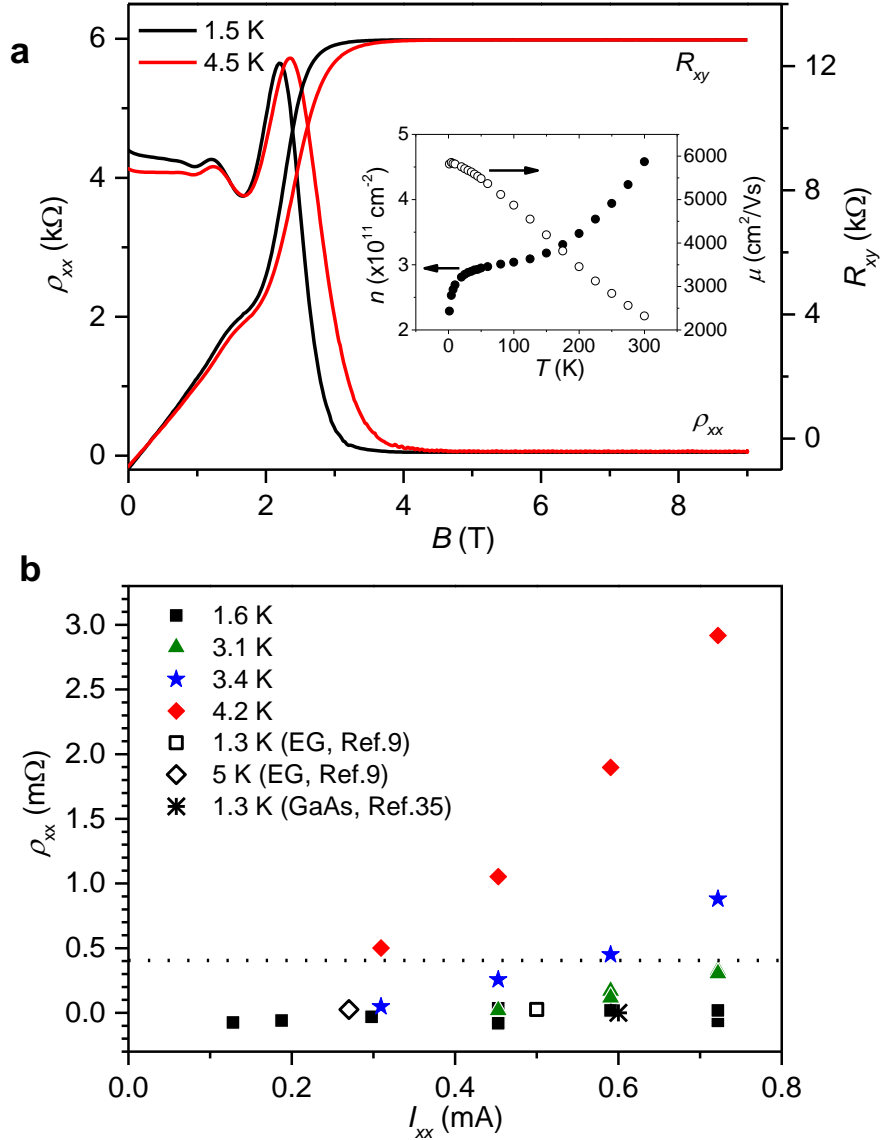

**Figure 4.** AC magnetotransport and DC precision measurements of $\rho_{xx}$ of sample A in a perpendicular magnetic field $B$. (a) Transport characteristics with $I_{xx} = 1$ µA, $n \approx 2.3\times10^{11}$ cm$^{-2}$, and $\mu \approx 5800$ cm$^2$/Vs for sample A. Inset: Graph of carrier density $n$ and mobility $\mu$ for temperatures from 1.5 K to 300 K. (b) Precision measurements of $\rho_{xx}$ at 9 T as a function of source-drain DC current at 1.6 K, 3.1 K, 3.4 K and 4.2 K. Dashed line at $\rho_{xx} = 0.4$ mΩ indicates the degree of quantization sufficient to produce $R_{xy}$ values within five parts in $10^9$ of the ideal quantized value, as described in the text. Results of earlier high-current studies of the QHE in graphene devices and in GaAs are included for comparison.

Near the onset of thermally-activated dissipation[10,36], a linear relationship is generally observed between the deviation of the Hall resistance $\Delta R_{xy}$ from the quantized value ($\frac{h}{2e^2}$) and the non-zero value of $\rho_{xx}$. A specialized two-terminal cryogenic current comparator (CCC) bridge[37] was employed to measure values of $R_{xy}$ at $T = 1.6$ K and $T = 4.2$ K against a precision 100 k$\Omega$ standard resistor at current levels of 0.3 mA, 0.45 mA, 0.6 mA and 0.72 mA. While GaAs-based QHE calibration of the 100 k$\Omega$ standard must be conducted at lower current levels and is thus less precise, the small differences measured at these two temperatures can be obtained with an uncertainty of better than $5 \times 10^{-9}$ in $R_{xy}$. Plotted against values of the longitudinal resistance $\rho_{xx}$ measured at 4.2 K for the same current levels, the deviation $\Delta R_{xy} = R_{xy}(4.2\ \text{K}) - R_{xy}(1.6\ \text{K})$ yields a slope $\Delta R_{xy}/\rho_{xx}(4.2\ \text{K}) \approx 0.164 \pm 0.01$ (see Supplement Fig. S5b). Thus, the accuracy of the QHE is maintained at the level of $5 \times 10^{-9}$ in sample A up to $I_{xx} \approx 0.72$ mA at $T = 3.1$ K and 9 T, where $\rho_{xx} \leq 0.4$ m$\Omega$ is measured (as noted by the dashed line in Fig. 4b).

In our samples the adsorbed molecular doping layer acts as a gate, and it is possible to control the carrier concentration through this effect. The concentration of adsorbed dopants was tuned after the initial fabrication of sample B, first by exposure to vapor from concentrated $HNO_3$, followed by gentle heating of the sample in vacuum[34,38]. AC transport measurements were made on sample B at seven levels of carrier density, as shown in Fig. 5a. Mobility in EG at low temperature is strongly dependent on carrier density, as demonstrated in sample B by the nearly inverse relationship between $\mu$ and $n$ for n-type carrier concentrations below $6.5 \times 10^{11}$ cm$^{-2}$. Similar levels of transport mobility near $\mu \approx 5800$ cm$^2$/Vs were obtained for sample A with $n \approx 2.3 \times 10^{11}$ cm$^{-2}$ and for sample B at carrier density $n_2 \approx 1.25 \times 10^{11}$ cm$^{-2}$.

Despite the presence of inhomogeneous strain and layer number in some areas of sample B, the measured mobility for $n_0 \approx 0.5 \times 10^{10}$ cm$^{-2}$ is $\mu \approx$ 43 700 cm$^2$/Vs, which is the highest reported mobility for any graphene sample of millimeter-scale dimensions. This confirms that high mobility can be maintained in EG in the presence of localized defects[20] if the density of these defects is not too high, and indicates that a 10 mm$^2$ area of sample B has very uniform carrier density at low temperature. For comparison, similar characteristics of $\mu$ and $n$ have been reported in gated, high-quality EG[23] devices of < 5 µm width. For those, the mobility was slightly lower at high carrier density and was seen to decrease for values of $n$ below $\approx 10^{11}$ cm$^{-2}$, an effect that we did not observe in the FTG sample B for much lower carrier densities. Exfoliated samples on h-BN[32] have exhibited mobility about twice what we measure, and the authors did not report any decrease in the mobility at low carrier density. More generally, the QHE in the millimeter-size EG samples described here far surpasses earlier results that we obtained using similar-scale CVD graphene[39].

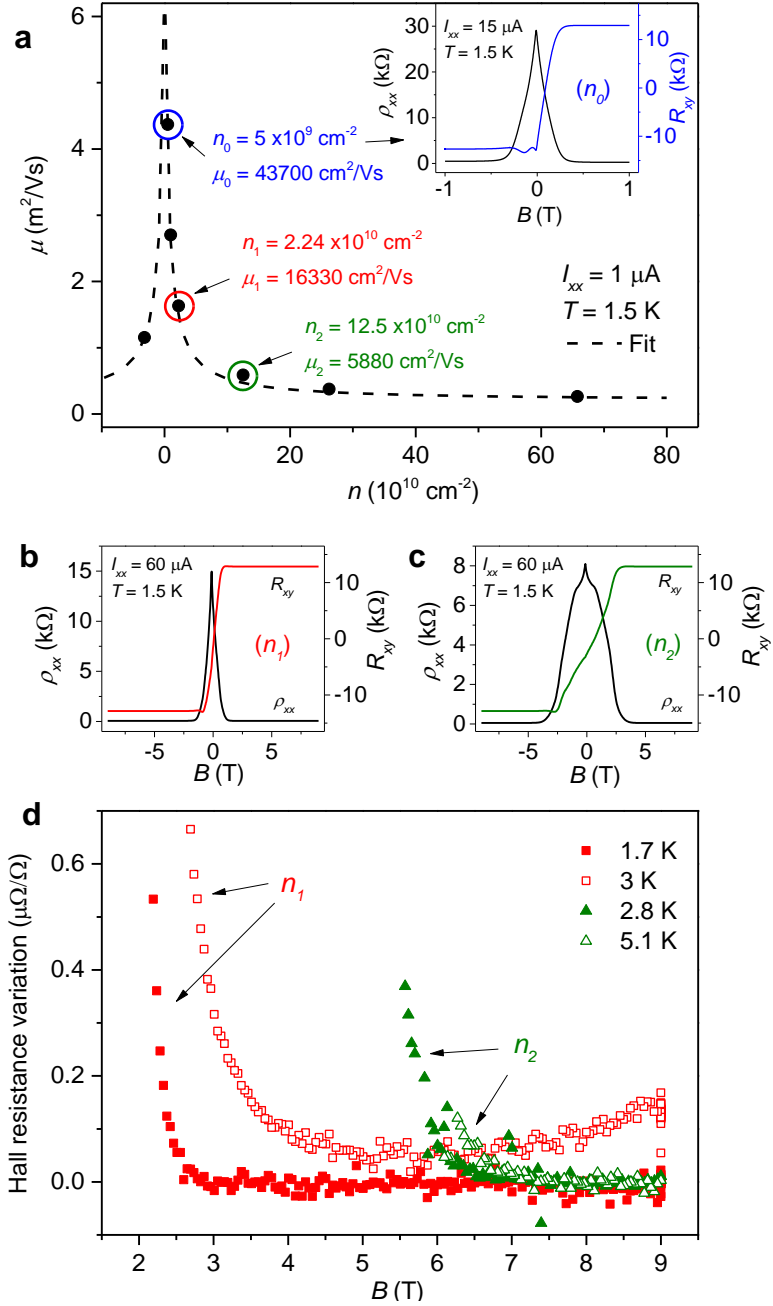

**Figure 5.** AC magnetotransport as a function of carrier density and DC precision QHE measurements of sample B. (a) Graph of mobility $\mu$ versus carrier density $n$ for sample B, calculated from the device conductivity $\sigma_{xx} = en\mu$ and the slope of the Hall resistance $R_{xy}(B)$. Starting at far left with low p-type doping ($3\times10^{10}$ cm$^{-2}$) changes in $n$ and $\mu$ were obtained by heating the sample at increasing temperatures of 320 °C – 340 °C in vacuum. Fitting of the data

to an inverse function $\mu = a + b\times(n-c)^{-1}$ for $n > 0$ results in the dotted curve, showing an approximate inverse relation, and this fit is mirrored about $n = 0$ to show that the same relation may exist for p-type carriers. Inset: transport characteristics for very low carrier density ($n_0$) identified by the blue circle. (b) Sample B transport characteristics for low carrier density ($n_1$). (c) Sample B transport characteristics for higher carrier density ($n_2$). (d) Precise measurements of the deviation in the Hall resistance $R_{xy}(B)$ plotted for the two levels of carrier density $n_1$ and $n_2$. Data for $n_1$ was taken at $I_{xx}$ = 19.4 µA, while data for $n_2$ was taken at $I_{xx}$ = 116 µA. The dependence on $n$ and $T$ is described in the text.

Precise values of $R_{xy}$ in sample B were measured at doping levels indicated by ($n_1$) and ($n_2$) shown in Fig. 5a, with magnetotransport data given in Figs. 5b,c. These measurements were made near the highest levels of current for which full quantization was maintained for each carrier density. At these current levels, $I_{xx}(n_1) \approx 0.0194$ mA and $I_{xx}(n_2) \approx 0.116$ mA, conventional CCC measurements of $R_{xy}$ could be based on standard resistors precisely calibrated against a NIST GaAs-based QHE standard. For the lower carrier density $n_1$ and $T \approx 1.7$ K, the value of $R_{xy}$ differed from $\frac{h}{2e^2}$ by less than the measurement uncertainty of $\pm 5 \times 10^{-9}$ within the range 3 T < $B$ < 9 T (see Fig. 5d). For similar conditions but with $T$ = 3.0 K, the value measured for $R_{xy}$ differed from the ideal value by about five parts in $10^8$ near $B$ = 6 T, with increased variation from the ideal value at lower and higher field. When sample B was tuned to higher carrier density ($n_2$), precise quantization was maintained over the range 7.5 T < $B$ < 9 T for higher measurement current of $I_{xx}$ = 0.116 mA and $T$ = 2.8 K. For the same range of $B$ and $I_{xx}$ = 0.116 mA, the value of $R_{xy}$ showed slight loss of quantization ($\Delta R_{xy}/R_{xy} \approx 1 \times 10^{-8}$) for $T$ = 5.1 K (Fig. 5d).

While our FTG graphene is produced on low-miscut SiC(0001) by annealing at 1900 °C in a near-equilibrium growth environment, improved QHE transport under relaxed conditions has also been reported in graphene which was grown on higher-miscut SiC(0001) by a hydrogen-supported CVD process[9], where step bunching of SiC appears to be strongly suppressed. Step-bunching and bilayer regions are sources of scattering, and our results suggest that one or both of these may contribute to non-uniform strain in the EG layer. From the low-field Hall slopes, we note that the extrapolated resistance of $\frac{h}{2e^2}$ occurs at $B \approx 4.8$ T for sample A and at $B \approx 2.6$ T for sample B. Relative to the extrapolated onset of the QHE regime, the less homogeneous sample B requires a much higher field to overcome dissipation and supports a much lower critical current at similar levels of magnetic field and temperature.

In conclusion, our observations for sample A clearly show that desirable magnetotransport properties are correlated with low and uniform strain in EG on SiC(0001) substrates. The uniform Raman G' band characteristics we observe compare favorably with those described in earlier reports[26-31] and provide the first example of EG with highly uniform strain at millimeter scale. The narrow G' FWHM values and their closely grouped distribution in sample A suggest reduced strain variation down to submicron scale[32]. In sample B, where bilayer inclusions are more common and strain is inhomogeneous in some areas, transport characteristics indicate that reduced scattering in the homogeneous EG regions allows efficient millimeter-scale transport. This is evident in sample B from the QHE results over an extended range of magnetic field and by mobility exceeding 40 000 cm$^2$/Vs for very low carrier density $n_0 \approx 1 \times 10^{10}$ cm$^{-2}$. Conditions favoring reduced strain in the EG lattice appear to support improved transport characteristics even in the presence of localized inhomogeneity. Thus, while a better understanding of strain inhomogeneity in monolayer EG is still needed, our results show that uniform lattice strain and reduced topographic variation contribute to improved 2D quantized conductance at elevated current and temperature, and this may provide direction for further advances in wafer-scale device fabrication.

## Methods

The samples were diced from two 76 mm SiC(0001) semi-insulating wafers (Cree, Inc.**) of nominal miscut 0.00°, with sample miscut measured to be ≤ 0.10° from AFM images. Samples were rinsed in HF and deionized water before processing, and arranged facing glassy pyrolytic graphite substrates (SPI Glas 22) with separation distance limited only by sample and substrate flatness. Processing was done in a graphite-lined resistive-element furnace (Materials Research Furnaces Inc.) with heating and cooling rates near 1.5 °C/s. The initial heating occurs in forming gas (96% Ar, 4% $H_2$) at 100 kPa with at least 30 min. cleaning of the substrates at 1050 °C, which may serve to hydrogenate the SiC surface[40,41]. The chamber was then flushed with Ar gas, and filled with 100 kPa Ar derived from 99.999% liquid Ar before annealing at 1900 °C, based on our earlier optimized processing results. The annealing process utilized a commercial process controller and a type-C thermocouple located a few cm above the sample.

Raman spectra were acquired under ambient conditions with a Renishaw InVia confocal Raman microscope equipped with 514.5 nm (2.41 eV) and 632.8 nm (1.96 eV) excitation lasers and an 1800 lines/mm grating while operating in 180° backscattering geometry. A 50× objective was used to focus the excitation laser light to an approximately 1 μm spot on the samples. Raman mapping measurements were performed using 514.5 nm excitation by raster scanning rectangular areas with a step size of 1 μm and collecting the Raman G' peak region with an exposure time of 10 s for each point. Raman maps were generated by fitting the spectra with a single Lorentzian peak and plotting the fitting parameters of FWHM and peak position at each pixel.

For initial transport characterization, four lock-in amplifiers monitored the longitudinal current $I_{xx}$ supplied at 13 Hz and three voltages developed in the device while we swept the perpendicular magnetic field strength $B$. We measured the Hall resistance $R_{xy} = V_{xy}/I_{xx}$ across the central pair of contacts. Longitudinal resistivity $\rho_{xx}$ was derived from the average resistance value measured across the other four symmetric contacts, scaled by the ratio of width to length ($w/L$ = 5.6 mm/1.8 mm) separating these terminals. The high precision longitudinal resistivity was measured between the central pair of contacts used to determine $R_{xy}$ and one set of adjacent contacts ($w/L$ = 5.6/0.9), using a nanovolt meter (EM Electronics model N11) and recorded automatically using an Agilent 3458A DMM. Periodically reversed current was supplied by a battery-powered ramping voltage source[37]. CCC measurements were made as described in Ref [37].

**Supporting Information**

Supporting Information includes (1) LEEM and Raman mapping of EG sample grown in Ar at 1630 °C. (2) Comparison of three EG processing methods at 1900 °C. (3) Enhanced optical imaging of FTG terrace structure. (4) DC precision measurements of sample A.

AUTHOR INFORMATION

**Corresponding Author**

*Email: yanfei.yang@nist.gov.

**Author Contributions**

Y.Y., G.C. and R.E.E. designed the experiments. Y.Y., G.C., C.C, C.-W.L., C.-I.L., P.M., I.G.C., R.M.F., G.R.J., R.M.F., and R.E.E. performed the experiments. Y.Y. and R.E.E. produced the samples and Y.Y fabricated the devices. Y.Y., G.C., A.R.H.W. and R.E.E. co-wrote the paper.

**Funding Sources**

The work of Y.Y. was supported by federal grant #70NANB12H185.

**Notes**

Identification of commercial products or services used in this work does not imply endorsement by the US government, nor does it imply that these products are the best available for the applications described.

ACKNOWLEDGMENT

The work of C.C, C.-W.L. and C.-I.L at NIST was made possible by arrangement with Prof. C.-T. Liang of National Taiwan University. We thank S. Lara-Avila of Chalmers University for fabricating the transport device made of graphene grown at low temperature.

ABBREVIATIONS

AFM, atomic force microscopy; EG, epitaxial graphene; FWHM, full width at half maximum; FTG, face-to-graphite; HF, hydrofluoric acid; QHE, quantum Hall effect; SiC, silicon carbide; LEEM, low-energy electron microscopy.

Supporting Information for

# Epitaxial graphene homogeneity and quantum Hall effect in millimeter-scale devices

*Yanfei Yang[†]\*, Guangjun Cheng[†], Patrick Mende[‡], Irene G. Calizo[†,§], Randall M. Feenstra[‡], Chiashain Chuang[†,Π], Chieh-Wen Liu[†,£], Chieh-I Liu[†,£], George R. Jones[†], Angela R. Hight Walker[†], and Randolph E. Elmquist[†]*

[†]National Institute of Standards and Technology (NIST), Gaithersburg, MD 20899-8171, USA

[‡]Department of Physics, Carnegie Mellon University, Pittsburgh, PA 15213-3890, USA

[§]Present address: Mechanical & Materials Engineering, Florida International University, Miami, FL 33174

[Π]Present address: Graduate School of Advanced Integration Science, Chiba University, Chiba, 263-8522, Japan

[£]Graduate Institute of Applied Physics, National Taiwan University, Taipei 106, Taiwan

\*E-mail: yanfei.yang@nist.gov

**Contents**

1. Comparison of three EG processing methods at 1900 °C

Figure S1 identifies significant differences in samples processed by three vapor confinement methods. Figure S1a shows an optical image of low-miscut Si-face SiC(0001) after processing face-to-face with another Si-face SiC(0001) sample at 1900 °C for 900 s. Very wide, irregular SiC terraces have been formed by step-bunching, with large step height between the terraces as shown in the AFM image of Fig. S1b. Figure S1c shows an AFM image of an open-to-Ar sample produced under the same conditions as sample A. Annealing at 1900 °C in the open configuration has resulted in the formation of small hexagonal pits in the substrate. In confocal optical images of sample A, as shown in Fig. S1d-e, the surface is uniform with only scattered inclusions of bilayer graphene (indicated by white arrows). Figure S4 compares sample A and sample B surface structure using contrast-enhanced optical images, which show the terrace steps as dark lines.

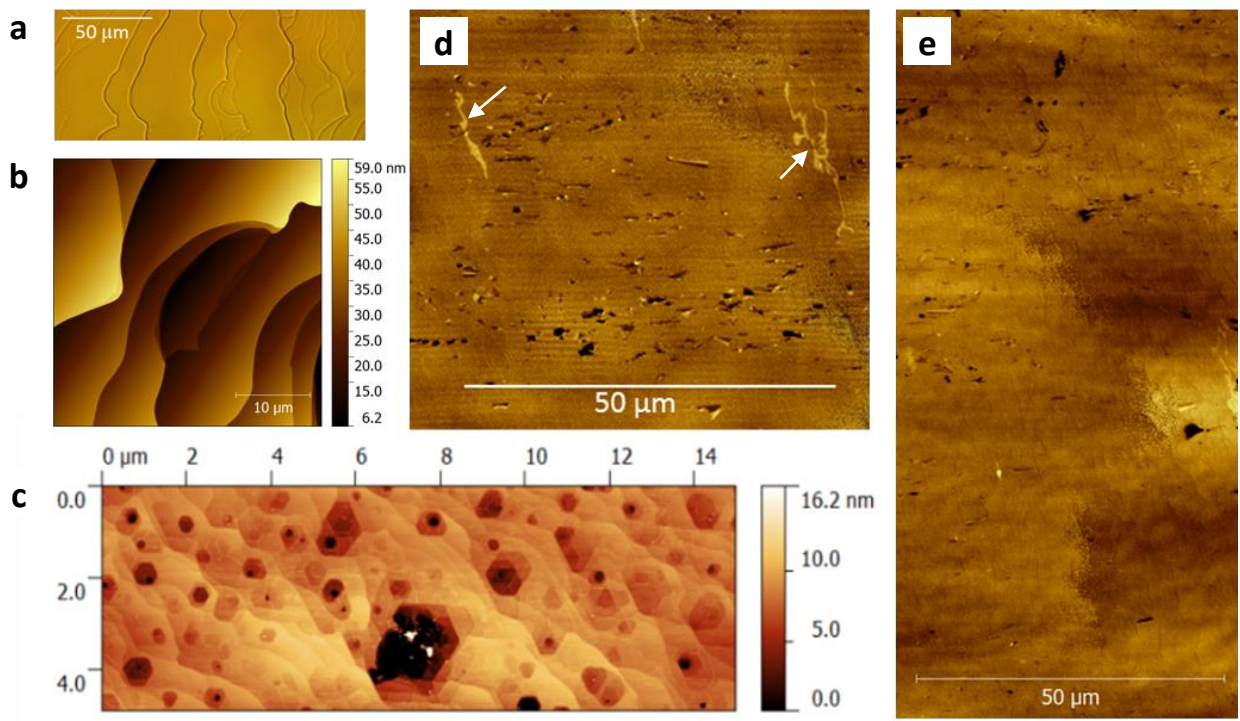

**Figure S1.** Comparison of three EG processing methods at the same temperature. (a) Optical image of a sample processed at 1900 °C for 900 s, arranged face-to-face with a second Si-face sample. (b) AFM image of a small region on the same face-to-face sample. (c) AFM image of an open-to-Ar sample annealed at 1900 °C for 210 s in the same processing run as Sample A, showing hexagonal pits formed in the SiC(0001) substrate. (d) Large confocal reflected light image of sample A, with irregularly shaped lighter patches of bilayer or multilayer graphene (indicated by the white arrows). Dark patches are dust, contamination from processing, or small patches of buffer layer graphene. (e) Another large confocal reflected light image of sample A. Terrace structure is faintly visible with mostly uniform EG. Lighter and darker shading over large areas on the right half are due to confocal imaging artefacts.

## 2. LEEM and Raman mapping of EG sample grown in Ar at 1630 °C.

An EG sample was grown in atmospheric-pressure Ar background gas at 1630 °C on a 6H-SiC(0001) substrate of average miscut $1.26° \pm 0.10°$. This produced strongly step-bunched terrace facets in the substrate. Overlapping LEEM and Raman maps were taken at regions identified by a grid of markings intentionally etched into the SiC substrate before EG growth. In the LEEM data shown in Figs. S2a-b, the labeled points can be identified from energy spectra (Fig. S2c) as monolayer EG, with monolayer EG covering the entire region of diameter 25 μm except within about 2 μm of the etched fiducial mark shown at upper right. LEEM imaging shows monolayer EG covering parallel terrace steps of 0.5 μm – 2 μm width; however, scattering from facet edges is not captured by LEEM, and delaminated bilayer EG could be present over ≈ 1.4% of the sample covered by those facets. The Raman maps shown in Figs. S2d-e cover a larger 50 μm diameter region, with a step size of 2 μm, and with an exposure time of 10 s for each point. Statistics derived from the Lorentzian fits show the G` FWHM ($63.8$ cm$^{-1}$ $\pm$ $2.6$ cm$^{-1}$) and position ($2747.4$ cm$^{-1}$ $\pm$ $1.7$ cm$^{-1}$). Essentially no large-scale variation in strain is observed at a level greater than the Raman spectral resolution of 2.0 cm$^{-1}$, indicating that strain in monolayer EG may be inhomogeneous at small length scales not resolved by Raman mapping.

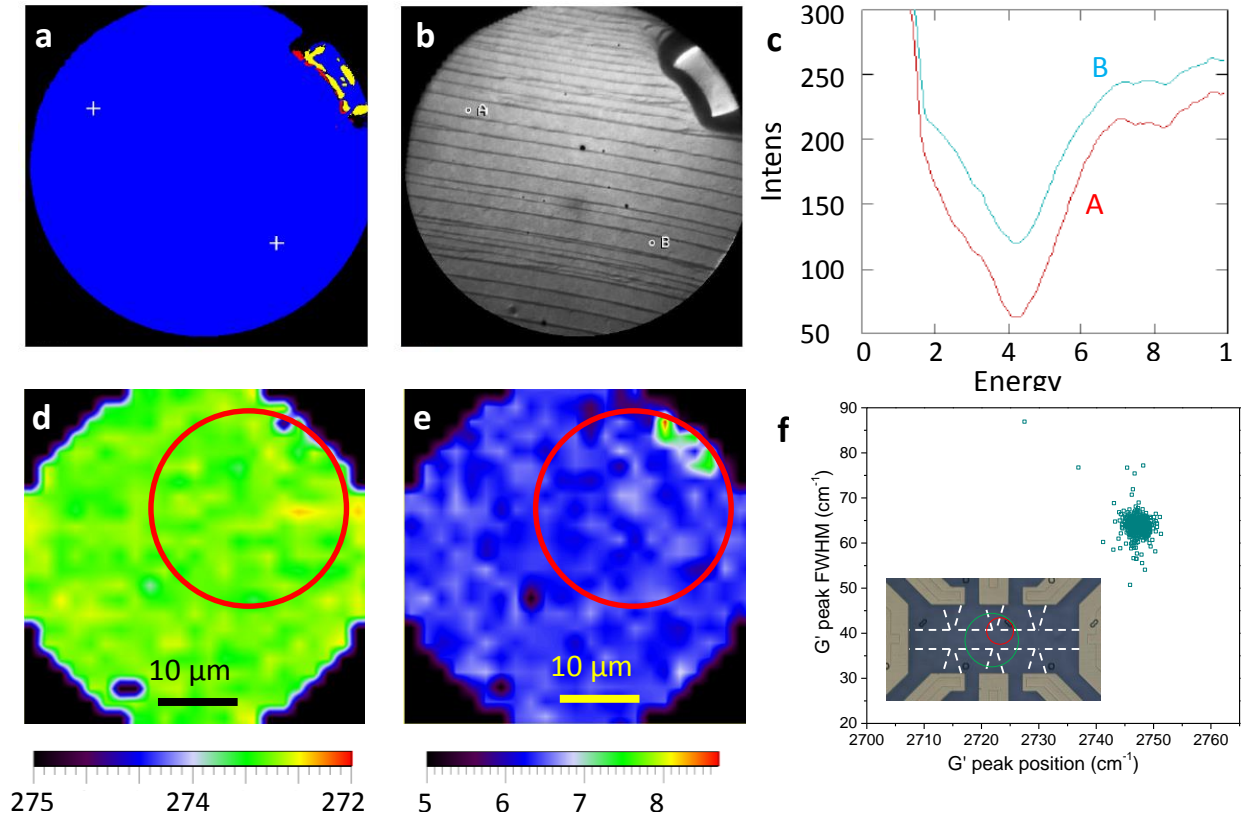

**Figure S2.** LEEM and Raman results for a region of an EG sample grown with the surface open to Ar background at 1630 °C. (a) LEEM map of EG thickness for a circular area of diameter ≈ 25 μm, from spectral analysis of layer number signatures (black = buffer, blue = monolayer, red = bilayer, yellow = trilayer). (b) Corresponding LEEM image acquired at electron energy of 6.0 eV. (c) LEEM energy spectra at two points marked in (a), each with a single intensity minimum characteristic of monolayer EG. (d) Micro-Raman image produced by mapping the G` FWHM over a circular area of diameter ≈ 44 μm, derived by fitting a single Lorentzian function to the G` band spectral region (2700 cm$^{-1}$ – 2800 cm$^{-1}$). (e) The centroid positions of the fits to the G` band. The red circles indicate the approximate region shown in the LEEM data. (f) Correlation between FWHM and peak position for G` peaks in (d-e). Inset: Optical image of a 20 μm wide Hall bar fabricated on this sample, where the green circles indicate the Raman area in (d-e) and the red circles indicate the LEEM area in (a-b).

3. **Typical Raman spectra for sample A.**

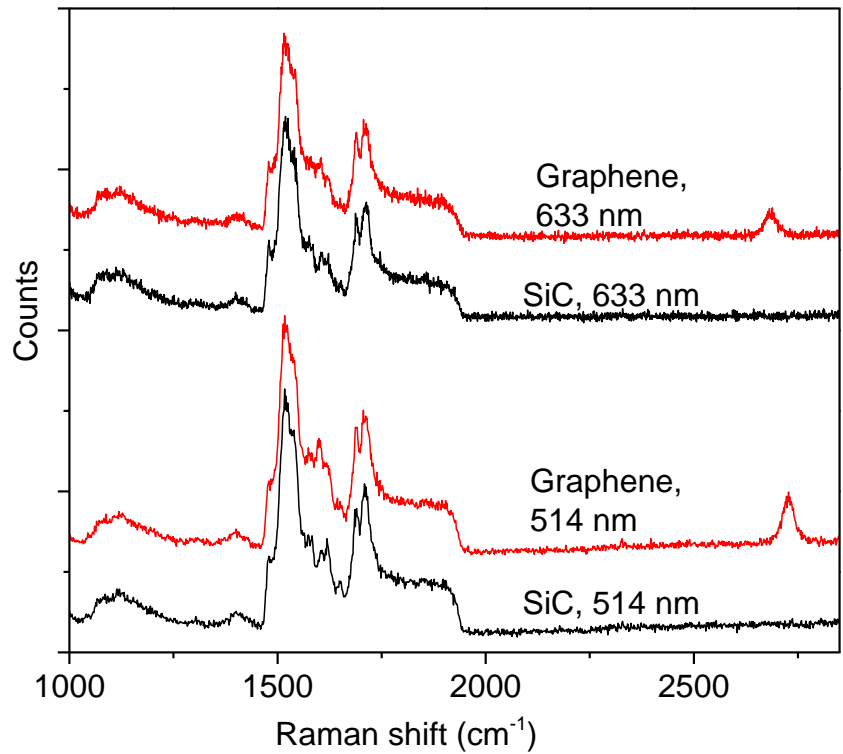

**Figure S3.** Raman spectra obtained on sample A with 633 nm and 514 nm lasers, compared to the spectra of bare SiC. Comparison shows no sign of the disorder-induced D peak ($\approx$ 1350 cm$^{-1}$).

4. **Enhanced optical imaging of FTG terrace structure.**

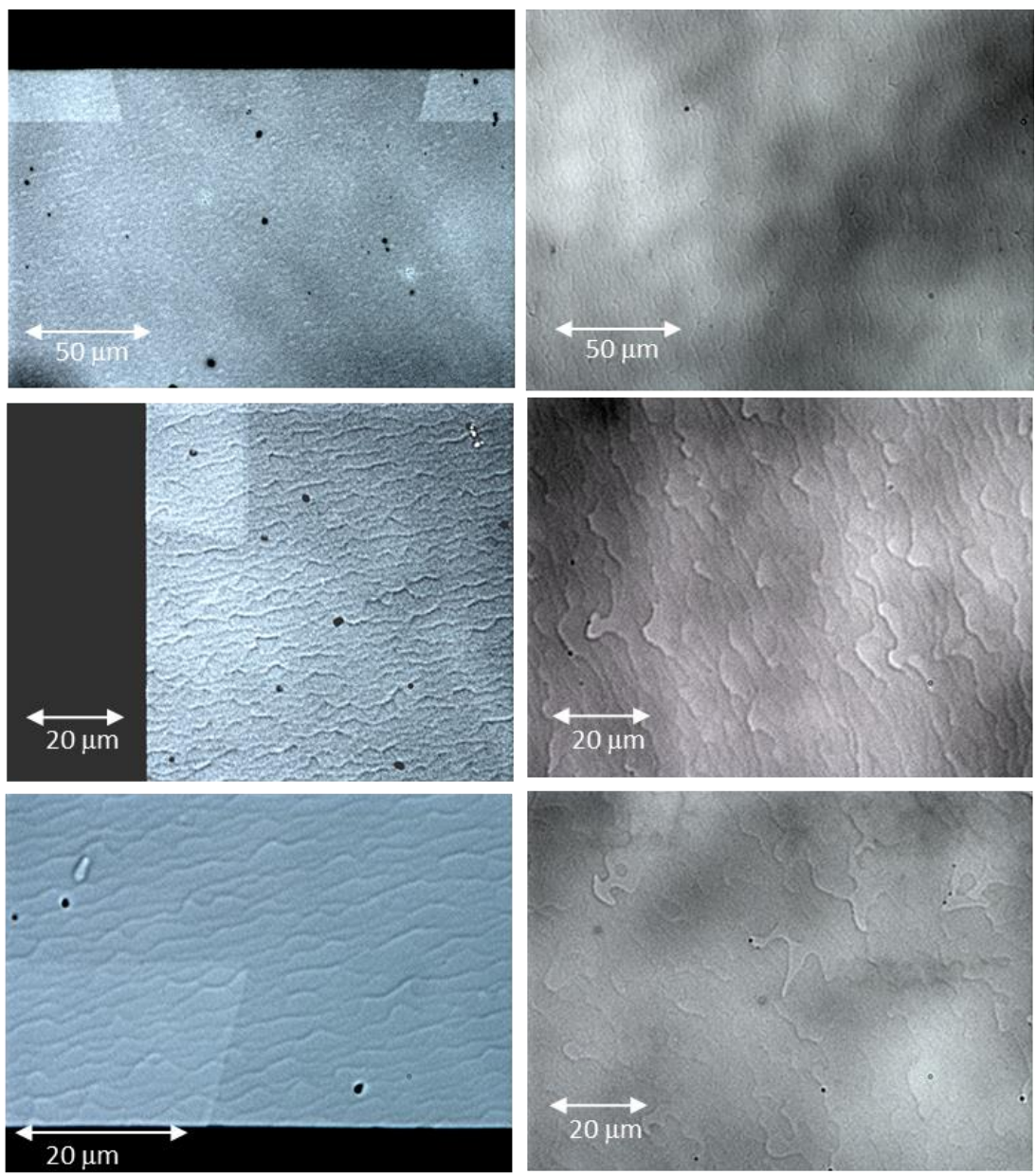

**Figure S4.** At left, three contrast-enhanced images from sample A taken near three electrical contacts using transmitted light. In these images the gold contact pads appear black and the regions where EG has been removed appear slightly lighter than the EG. The terraces on sample A show relatively little variation in size and shape. The three images at right show regions of sample B. At top right, the imaged region has small terraces similar to those on sample A. The lower two images show regions where step bunching is more advanced, with the wider terraces having irregular shape. Darker shadowing is due to the imaging process using transmitted light.

## 5. DC precision measurements of sample A.

Precision measurements of the longitudinal resistivity $\rho_{xx}$ were made using a DC nanovoltmeter (EM Electronics model N11) with a battery power supply. The $V_{xx}$ terminals were arranged in a rectangle spanning the device, with an area of $w = 5.6$ mm by $L = 0.9$ mm. Two of these terminals were used to measure $V_{xy}$, employing a cryogenic current comparator with maximum bridge output voltage of 10 V.

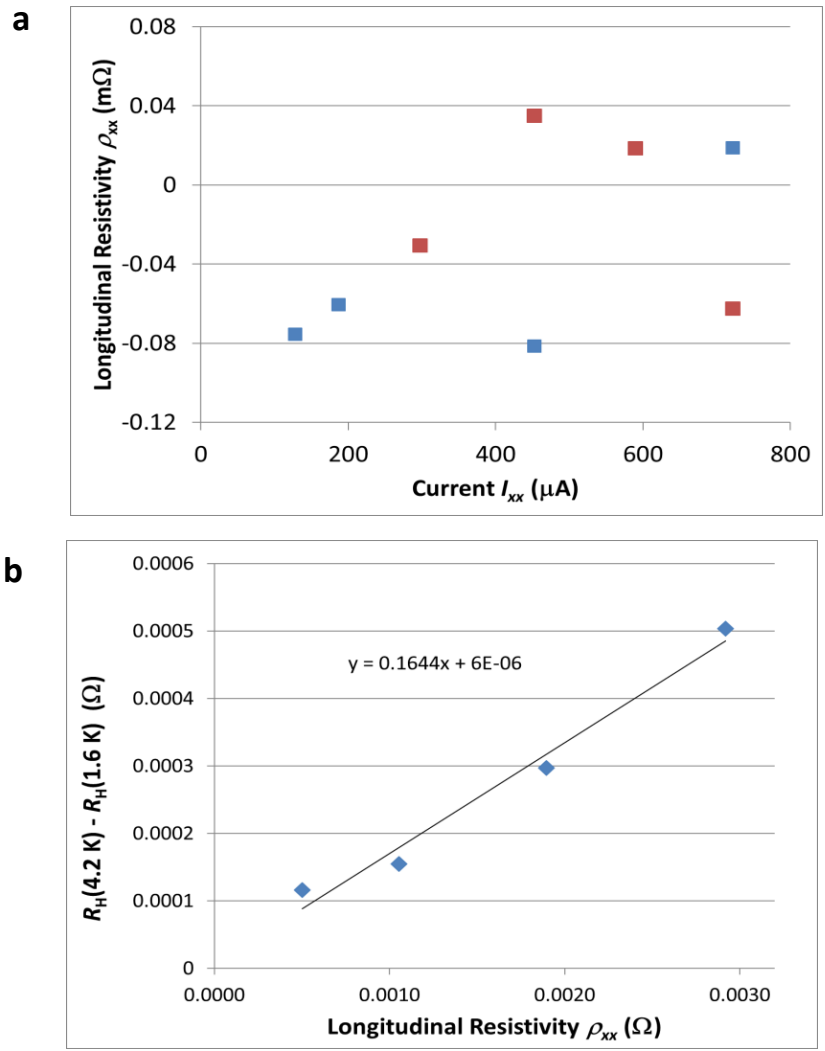

**Figure S5.** DC precision measurement of sample A. (a) Longitudinal resistivity $\rho_{xx}(1.6\ \text{K})$ as a function of current, magnified from Fig. 4b of the main paper. Marker colors indicate

measurements made on two different days. The standard deviation of the data points varies from 0.15 m$\Omega$ near 200 µA to 0.04 m$\Omega$ near 720 µA. (b) Relative resistance change $\Delta R_H = \{R_{xy}(1.6\ K) - R_{xy}(4.2\ K)\}/R_{xy}(1.6\ K)$ as a function of longitudinal resistivity at 4.2 K. The standard deviation of the data points is 0.01 µ$\Omega$/$\Omega$ or lower.

**Table S1.** Statistics of the Raman G' band for sample A.

| Sample A | Points | Mean position (cm$^{-1}$) | Position Std (cm$^{-1}$) | Mean Width (cm$^{-1}$) | Width Std (cm$^{-1}$) |
|---|---|---|---|---|---|
| Area 1 | 442 | 2728 | 2.5 | 31.6 | 1.72 |
| Area 2 | 272 | 2730 | 3.0 | 31.7 | 2.57 |
| Area 3 | 480 | 2730 | 3.3 | 32.3 | 3.80 |

**Table S2.** Statistics of the Raman G` band for sample B.

| Sample B | Points | Mean position (cm$^{-1}$) | Position Std (cm$^{-1}$) | Mean Width (cm$^{-1}$) | Width Std (cm$^{-1}$) |
|---|---|---|---|---|---|
| Area 1 | 1058 | 2735 | 4.3 | 34.7 | 4.6 |
| Area 2 | 660 | 2738 | 6.3 | 37.3 | 5.18 |
| Area 3 | 63 | 2738 | 3.0 | 36.8 | 3.18 |
| Area 4 | 42 | 2735 | 3.2 | 38.6 | 5.24 |
| Area 5 | 49 | 2735 | 5.4 | 34.9 | 3.19 |
| Area 6 | 63 | 2734 | 4.6 | 37.3 | 5.51 |
| Area 7 | 49 | 2739 | 7.5 | 52.1 | 8.43 |
| Area 8 | 56 | 2746 | 6.4 | 43.6 | 8.93 |
| Area 9 | 36 | 2753 | 18.2 | 70.5 | 29.68 |
| Area 10 | 56 | 2754 | 23.0 | 56.7 | 29.84 |